\documentclass{ws-ijmpd}
\usepackage[super,compress]{cite}
\begin{document}

\markboth{Yu Li}
{Dynamics of two-scalar-field cosmological models beyond the exponential potential}

\catchline{}{}{}{}{}
%
%%%%%%%%%%%%%%%%%%%%%%%%%%%%%%%%%%%%%%%%%%%%%%%%%%%%%%%%%%%%%%%%%%%%

\title{Dynamics of two-scalar-field cosmological models beyond the exponential potential}

\author{Yu Li}

\address{Department of Physics, Dalian Maritime University, Dalian 116026, China\\
leeyudlmu@gmail.com}

%\author{SECOND AUTHOR}
%
%\address{Group, Laboratory, Address\\
%City, State ZIP/Zone, Country\\
%second\_author@group.com}

\maketitle

\begin{history}
\received{Day Month Year}
\revised{Day Month Year}
\end{history}

\begin{abstract}
In this paper, we discuss the dynamics of two- scalar-field cosmological models. Unlike in the situation of 
exponential potential, we find that there are late time attractors in which one scalar field dominates
the energy density of universe and the other one decay. We also discuss the possibility of multiple attractors model which is useful to realize the evolve of universe from a scaling era to recent acceleration era. We also give the conditions of the existence of multiple attractors.
\end{abstract}

\keywords{Dynamics of cosmological model; Mutli-scalar field; Beyond exponential potential}

\ccode{PACS numbers:}98.80.-k, 98.80.Jk,95.35.+d

%\tableofcontents
\section{Introduction}
Observational evidence for the universe suggest that the universe is in an accelerating phase of expansion\cite{r10,r11}. This implies the existence of a negative dark energy. The simplest dark energy
is a positive cosmological constant. However, the prediction for cosmological constant in quantum field theory differs from the measured value by many orders of magnitude \cite{r12}, this is so-called cosmological constant problem \cite{r13,r14}. One method to solve this issue is dynamical dark energy, i.e. dark energy can evolve slowly with the cosmic expansion. This could be achieved through scalar field models \cite{r15,r16,b3}.

One important scalar field model is a scalar field with an  expanential potential energy. It play 
an important role in higher order or higher dimensional gravity theories, and the dynamics of these models can be easily discussed by the method of autonomous system \cite{r17,r18,r19}. However, the exponential form of potential energy is not the only possibility, so it is necessary to extend the discussion to a more general potential. The dynamics of scalar field model beyond the 
expanential potential have studied in\cite{r3,r4}.

For single scalar field model, if the potential is too steep, the universe will not accelerate at late time\cite{r1}. While, if the multi-scalar fields model is considered, the situation will be different.
The universe with multi-scalar fields can evolve to accelerate era even if individual scalar field is 
too steep to drive the accelerating of the universe, i.e. assisted inflation\cite{r5,r20,r21,r22} and assisted quintessence \cite{r23,r24,r25}. One may realize that the potentials
investigated in these papers are of the exponential form. In this paper, we will extend the idea to 
arbitrary potential.

We first rewirte the evolution equaions of universe with two scalar fields to the dynamical autonomous system in Sec.\ref{s2}. And then we give all the fixed points of this system in Sec.\ref{s3}. 
In this paper, we 
only focus on stabble fixed points. We find that there are single scalar dominating solutions under 
some conditions. In Sec.\ref{s4}, we discuss the issue of multiple attractors which is useful to  realize the evolve of universe from a scaling era to recent
acceleration era. We also give the conditions of the existence of multiple attractors. The summary is in Sec.\ref{s5}

\section{ Dynamical  Autonomous System\label{s2}}
To start, we consider a spatially flat Friedmann-Robertson-Walker universe containing two scalar fields $\phi_i,(i=1,2)$ with the potentials $V_i(\phi_i)$ and a perfect fluid with state equation $p_{\gamma}=(\gamma-1)\rho_{\gamma}$, where $\gamma$ is an adiabatic index and we shall assume that $0<\gamma<2$.

The Einstein field equations, yield the Friedmann's equation and Raychduri's equation:
\begin{eqnarray}
H^2&=&\frac{\kappa}{3}(\rho_{\gamma}+\sum_i\rho_{\phi_i})\label{e1}\\
\dot{H}&=&-\frac{\kappa}{2}(\rho_{\gamma}+p_{\gamma}+\sum_i\rho_{\phi_i}+\sum_ip_{\phi_i})\label{e2}
\end{eqnarray}
where $\kappa\equiv 8\pi G$ and a overdot represents differentiation with respect to cosmic time. In this paper, we choose the units that $\kappa=c=1$.

The energy density and the pressure for scalar field $\phi_i$ are:
\begin{equation}
\label{e3}
\rho_{\phi_i}=\frac{1}{2}\dot{\phi}_i+V_i(\phi_i),~~~~p_{\phi_i}=\frac{1}{2}\dot{\phi}_i-V_i(\phi_i)
\end{equation}

The conservation equation for perfect fluid and the Klein-Gordon equations for scalar fields are:
\begin{eqnarray}
&&\dot{\rho}_{\gamma}+3\gamma H\rho_{\gamma}=0\label{e4}\\
&&\ddot{\phi}_i+3H\dot{\phi}_i+V_i'=0\label{e5}
\end{eqnarray}
where $V_i'\equiv d V_i/d \phi_i$.

Eqs. (\ref{e1}), (\ref{e4}) and (\ref{e5}) or  Eqs.(\ref{e2}), (\ref{e4}) and (\ref{e5}) characterize a closed system which can determine the cosmic behavior. To analyze the dynamical behavior of the 
universe, we can introduce the following dimensionless variables as in\cite{r1,r2}:
\begin{equation}
\label{e6}
\Omega^2\equiv\frac{\rho_{\gamma}}{3H^2},~~x_i^2\equiv\frac{\phi_i^2}{6H^2},~~y_i^2\equiv\frac{V_i}{3H^2},~~\lambda_i\equiv-\frac{V_i'}{V_i}
\end{equation}

Using the new variables in (\ref{e6}), one can rewrite the Friedmann equation (\ref{e1}) as:
\begin{equation}
\label{e8}
1=\Omega^2+\sum_i x_i^2+\sum_i y_i^2
\end{equation}
and the Eqs (\ref{e2}), (\ref{e4}) and (\ref{e5}) can be rewritten in following dynamical form:
\begin{eqnarray}
\frac{dx_i}{dN}&=&-3x_i+\frac{\sqrt{6}}{2}\lambda_i y_i^2+\frac{3}{2}x_i\left[(2-\gamma)\sum_ix_i^2+
\gamma\left(1-\sum_iy_i^2\right)\right]\label{e9}\\
\frac{dy_i}{dN}&=&-\frac{\sqrt{6}}{2}\lambda_ix_iy_i+\frac{3}{2}y_i\left[(2-\gamma)\sum_ix_i^2+
\gamma\left(1-\sum_iy_i^2\right)\right]\label{e10}\\
\frac{d\lambda_i}{dN}&=&-\sqrt{6}\lambda_i^2(\Gamma_i-1)x_i\label{e11}
\end{eqnarray}
where $N\equiv \ln a$ and 
\begin{equation}
\label{e12}
\Gamma_i\equiv\frac{V_iV_i''}{V_i'^2}
\end{equation}

In order to build the Eqs.  (\ref{e9}-\ref{e11}) into autonomous system, one can consider $\Gamma_i$ as a function of $\lambda_i$ like in \cite{r3,r4}, i.e.:
\begin{equation}
\label{e13}
\Gamma_i(\lambda_i)=f_i(\lambda_i)+1
\end{equation}
and Eq.(\ref{e11}) becomes:
\begin{equation}
\label{e14}
\frac{d\lambda_i}{dN}=-\sqrt{6}\lambda_i^2f_i(\lambda_i)x_i
\end{equation}
Equations (\ref{e9},\ref{e10},\ref{e14}) definitely describe a dynamical  autonomous system.

Eq.(\ref{e13})  can be seen as a  restriction on the  form of the potential. In fact, many quintessential potentials satisfy these constraints\cite{r3,r4,r16}.
For exponential potentials, one can obtain that $f_i(\lambda_i)=0$. 
In this case The $3i$-dimensional autonomous system reduces to $2i$-dimensional autonomous system.
For phantom potential $V_i(\phi_i)=\frac{V_0}{[\cosh(\sigma\phi_i)]^n}$, 
one can obtain that $f_i(\lambda_i)=\frac{1}{n}-\frac{n\sigma^2}{\lambda_i^2}$\cite{r3}.

On the other hand, one can also choose the form of $f(\lambda)$ or $\Gamma(\lambda)$, and then derive the form of potentials reversely. 
One approach to obtain the potential can be found in \cite{r3} as follow:

Since the potential $V(\phi)$ is only the function of the field $\phi$, by the definition (\ref{e6},\ref{e12}),  $\lambda$ and $\Gamma$ can
be written as:
\begin{equation}
\label{eb1}
\lambda=P(\phi)~~~~~~\Gamma=Q(\phi)
\end{equation}
If the inverse function of $P(\phi)$ exists, then we have
\begin{equation}
\label{eb2}
\Gamma=Q(P^{-1}(\lambda))\equiv\mathcal{F}(\lambda)
\end{equation}
According to the definitions of $\lambda$ and $\Gamma$, Eq.(\ref{eb2}) can be rewritten as 
\begin{equation}
	\label{eb3}
	V''=\frac{V'^2}{V}\mathcal{F}\left(-\frac{V'}{V}\right)\equiv F(V,V')
\end{equation}
Let $h\equiv V'$ then
\begin{equation}
\label{eb4}
\frac{dh}{dV}=\frac{1}{h}F(V,h)=\frac{h}{V}\mathcal{F}\left(-\frac{h}{V}\right)
\end{equation}
Having figured out $h(V)$, one can solve $V'(\phi)=h(V(\phi))$ to obtain the potential $V(\phi)$.

For example, if one choose $\Gamma=1+\frac{\alpha}{\lambda^2}$ as in \cite{r3}, we can solve to 
the potential $V(\phi)=V_0\exp[\frac{\alpha}{2}\phi(\phi+\beta)]$. 
Another interesting cosmological model where the universe can evolve from a scaling attractor to a de-Sitter-like attractor 
discussed in  \cite{r3} is $\Gamma=1+\frac{1}{\beta}+\frac{\alpha}{\lambda}$, which corresponds to $V(\phi)=\frac{V_0}{(\eta+\exp(-\alpha\phi))^\beta}$.
Another approach to obtaing the potential can be found in \cite{r4}.
\section{Fixed Points and Stability\label{s3}}
For simplicity, we define $dx_i/dN\equiv X_i(x_i,y_i,\lambda_i)$, $dy_i/dN\equiv Y_i(x_i,y_i,\lambda_i)$, $d\lambda_i/dN\equiv\Lambda_i(x_i,y_i,\lambda_i)$. 
The fixed points can be obtained by set $X_i=Y_i=\Lambda_i=0$.

From Eq.(\ref{e14}), it is straightforward to see that $x_i=0$, $f_i(\lambda_i)=0$ or $\lambda_i=0$ can 
make $\Lambda_i=0$. However, as argured in \cite{r4}, $\lambda_i=0$ is not the  sufficient condition for $\lambda_i^2f_i(\lambda_i)=0$. 

For example, the phantom potential $V_i(\phi_i)=\frac{V_0}{[\cosh(\sigma\phi_i)]^n}$corresponds to $f_i(\lambda_i)=\frac{1}{n}-\frac{n\sigma^2}{\lambda_i^2}$, 
which means that $\lambda_i^2f_i(\lambda_i)=\frac{1}{n}\lambda_i^2-n\sigma^2\neq 0$ even if $\lambda_i=0$.

Therefore, the  sufficient conditions for $\Lambda_i=0$ are $x_i=0$,  $f_i(\lambda_i)=0$ or $\lambda_i^2f_i(\lambda_i)=0$. 

\begin{table}
	\tbl{The fixed points of autonomous system described by Eqs. (\ref{e9},\ref{e10},\ref{e14}).}
	{\begin{tabular}{c|cccccc|c||c|cccccc|c}\toprule
			Point&$\lambda_{1c}$&$x_{1c}$&$y_{1c}$&$\lambda_{2c}$&$x_{2c}$&$y_{2c}$&Existence&Point&$\lambda_{1c}$&$x_{1c}$&$y_{1c}$&$\lambda_{2c}$&$x_{2c}$&$y_{2c}$&Existence\\
\colrule
		$A_0$&$0$&$0$&$0$&$0$&$0$&$0$&for all $\gamma$&	$\bar{C}_2$&$\lambda_a$&$0$&$0$&$0$&$0$&$1$&for all $\gamma$\\
		$A_1$&$0$&$\cos\theta$&$0$&$0$&$\sin\theta$&$0$&for all $\gamma$&$D_0$&$\lambda_{a1}$&$0$&$0$&$\lambda_{a2}$&$0$&$0$&for all $\gamma$\\	
		$A_2$&$0$&$0$&$\psi$&$0$&$0$&$\psi'$&for all $\gamma$&$E_0$&$\lambda_1^*$&$0$&$0$&$\lambda_{a}$&$0$&$0$&for all $\gamma$\\
		$B_0$&$0$&$0$&$0$&$\lambda_2^*$&$0$&$0$&for all $\gamma$&$\bar{E}_0$&$\lambda_{a}$&$0$&$0$&$\lambda_2^*$&$0$&$0$&for all $\gamma$\\
		$\bar{B}_0$&$\lambda_1^*$&$0$&$0$&$0$&$0$&$0$&for all $\gamma$&$E_{\pm1}$&$\lambda_1^*$&$\pm1$&$0$&$\lambda_a$&$0$&$0$&for all $\gamma$\\
		$B_1$&$0$&$\cos\theta$&$0$&$\lambda_2^*$&$\sin\theta$&$0$&for all $\gamma$&$\bar{E}_{\pm1}$&$\lambda_{a}$&$0$&$0$&$\lambda_2^*$&$\pm1$&$0$&for all $\gamma$\\
		$\bar{B}_1$&$\lambda_1^*$&$\sin\theta$&$0$&$0$&$\cos\theta$&$0$&for all $\gamma$&$E_2$&$\lambda_1^*$&$\frac{\lambda_1^*}{\sqrt{6}}$&$G_1$&$\lambda_{a}$&$0$&$0$&$|\lambda_1^*|<\sqrt{6}$\\
		$B_2$&$0$&$0$&$1$&$\lambda_2^*$&$0$&$0$&for all $\gamma$&$\bar{E}_2$&$\lambda_{a}$&$0$&$0$&$\lambda_2^*$&$\frac{\lambda_2^*}{\sqrt{6}}$&$G_2$&$|\lambda_2^*|<\sqrt{6}$\\
		$\bar{B}_0$&$\lambda_1^*$&$0$&$0$&$0$&$0$&$1$&for all $\gamma$&$E_3$&$\lambda_1^*$&$\frac{\sqrt{6}\gamma}{2\lambda_1^*}$&$P_1$&$\lambda_{a}$&$0$&$0$&$\lambda_1^{*2}>3\gamma$\\
		$B_3$&$0$&$0$&$0$&$\lambda_2^*$&$\frac{\lambda_2^*}{\sqrt{6}}$&$G_2$&$|\lambda_2^*|<\sqrt{6}$&$\bar{E}_3$&$\lambda_{a}$&$0$&$0$&$\lambda_2^*$&$\frac{\sqrt{6}\gamma}{2\lambda_2^*}$&$P_2$&$\lambda_2^{*2}>3\gamma$\\
		$\bar{B}_3$&$\lambda_1^*$&$\frac{\lambda_1^*}{\sqrt{6}}$&$G_1$&$0$&$0$&$0$&$|\lambda_1^*|<\sqrt{6}$&$F_0$&$\lambda_1^*$&$0$&$0$&$\lambda_2^*$&$0$&$0$&for all $\gamma$\\
		$B_4$&$0$&$0$&$0$&$\lambda_2^*$&$\frac{\sqrt{6}\gamma}{2\lambda_2^*}$&$P_2$&$\lambda_2^{*2}>3\gamma$&$F_1$&$\lambda_1^*$&$\cos\theta$&$0$&$\lambda_2^*$&$\sin\theta$&$0$&for all $\gamma$\\
		$\bar{B}_4$&$\lambda_1^*$&$\frac{\sqrt{6}\gamma}{2\lambda_1^*}$&$P_1$&$0$&$0$&$0$&$\lambda_1^{*2}>3\gamma$&$F_2$&$\lambda_1^*$&$\frac{\lambda_1^*}{\sqrt{6}}$&$G_1$&$\lambda_2^*$&$0$&$0$&$|\lambda_1^*|<\sqrt{6}$\\
		$C_0$&$0$&$0$&$0$&$\lambda_a$&$0$&$0$&for all $\gamma$&$F_3$&$\lambda_1^*$&$\frac{\sqrt{6}\gamma}{2\lambda_1^*}$&$P_1$&$\lambda_2^*$&$0$&$0$& $\lambda_1^{*2}>3\gamma$\\
		$\bar{C}_0$&$\lambda_a$&$0$&$0$&$0$&$0$&$0$&for all $\gamma$&$F_4$&$\lambda_1^*$&$0$&$0$&$\lambda_2^*$&$\frac{\lambda_2^*}{\sqrt{6}}$&$G_2$&$|\lambda_2^*|<\sqrt{6}$\\
		$C_{\pm1}$&$0$&$\pm1$&$0$&$\lambda_a$&$0$&$0$&for all $\gamma$&$F_5$&$\lambda_1^*$&$0$&$0$&$\lambda_2^*$&$\frac{\sqrt{6}\gamma}{2\lambda_2^*}$&$P_2$& $\lambda_2^{*2}>3\gamma$\\
		$\bar{C}_{\pm1}$&$\lambda_a$&$0$&$0$&$0$&$\pm1$&$0$&for all $\gamma$&$F_6$&$\lambda_1^*$&$\frac{\lambda_{eff}^2}{\sqrt{6}\lambda_1^*}$&$Q_1$&$\lambda_2^*$&$\frac{\lambda_{eff}^2}{\sqrt{6}\lambda_2^*}$&$Q_2$& $|\lambda_{eff}|<\sqrt{6}$\\
		$C_2$&$0$&$0$&$1$&$\lambda_a$&$0$&$0$&for all $\gamma$&$F_7$&$\lambda_1^*$&$\frac{\sqrt{6}\gamma}{2\lambda_1^*}$&$P_1$&$\lambda_2^*$&$\frac{\sqrt{6}\gamma}{2\lambda_2^*}$&$P_2$&$\lambda_{eff}^{2}>3\gamma$\\
		\botrule
	\end{tabular} \label{t1}}
\end{table}

The fixed points of autonomous system described by Eqs. (\ref{e9},\ref{e10},\ref{e14}) are shown in Table\ref{t1}. 
where $\lambda_1^*$ and $\lambda_2^*$ are the value of $\lambda$ which make $f_1(\lambda_1)=0$ and $f_2(\lambda_2)=0$ respectively, i.e. they are the zero points of $f_1(\lambda_1)$ and $f_2(\lambda_2)$. $\lambda_a$
means an arbitrary value and 
\begin{eqnarray}
&&0\leq\psi\leq1,~~\psi'\equiv\sqrt{1-\psi^2}\nonumber\\
&&G_1\equiv\sqrt{1-\frac{\lambda_{1}^{*2}}{6}},~~G_2\equiv \sqrt{1-\frac{\lambda_{2}^{*2}}{6}}\nonumber\\
&&P_1\equiv\frac{\sqrt{6\gamma(2-\gamma)}}{2\lambda_1^*},~~P_2\equiv\frac{\sqrt{6\gamma(2-\gamma)}}{2\lambda_2^*}\nonumber\\
&&Q_1\equiv\frac{\lambda_{eff}}{\lambda_1^*}\sqrt{1-\frac{\lambda_{eff}^{2}}{6}},~~Q_2\equiv\frac{\lambda_{eff}}{\lambda_2^*}\sqrt{1-\frac{\lambda_{eff}^{2}}{6}}\label{e15}
\end{eqnarray}
where
\begin{equation}
\label{e16}
\frac{1}{\lambda_{eff}^2}\equiv\frac{1}{\lambda_1^{*2}}+\frac{1}{\lambda_2^{*2}}
\end{equation}

The properties of  each fixed points are determined by the eigenvalues of the Jacobi matrix.
It is found that there are three fixed points can be stabble under some conditions:
the de Sitter-like solution $A_2$, the assisted scalar field dominated
solution case $F_6$ and the assisted scaling solution $F_7$.

%\section{Single Scalar Field  Attractors}
Strictly speaking, $A_2$ is not a single point in six-dimensional phase space. It is a quadrant of a circle in $y_1-y_2$ subspace for $0\leq\psi\leq1$.

The eigenvalues of the Jacobi matrix for $A_2$ are $\{0,0,0,-3,-3,-3\gamma\}$, which means it is a 
nonhyperbolic point. The stability of this point cannot be determined by the linearization method simply. 

To obtain the stability properties of $A_2$, we take the coordinate transformation: $y_1\rightarrow\bar{y}_1=y_1-\psi$, $y_2\rightarrow\bar{y}_2=y_2-\psi'$.
Then the corrdinate of $A_2$ transfer to $A_2'(0,0,0,0,0,0)$. One can transfer the Eqs. (\ref{e9},\ref{e10},\ref{e14}) to  $dx_i/dN\equiv X_i(x_i,\bar{y}_i,\lambda_i)$, $d\bar{y}_i/dN\equiv \bar{Y}_i(x_i,\bar{y}_i,\lambda_i)$, $d\lambda_i/dN\equiv\Lambda_i(x_i,\bar{y}_i,\lambda_i)$. When $\psi\neq 0,1$, the eigensystem of Jacobi matrix $J$ for $A_2'$ are:
\begin{eqnarray}
&&\{-3,[1,0,0,0,0,0]\},~~\{-3,[0,1,0,0,0,0]\},\nonumber\\
&&\left\{-3\gamma,\left[0,0,\frac{\psi}{\psi'},1,0,0\right]\right\},~~\left\{0,\left[0,0,-\frac{\psi'}{\psi},1,0,0\right]\right\},\nonumber\\
&&\left\{0,\left[\frac{\psi^2}{\sqrt{6}},0,0,0,1,0\right]\right\},~~\left\{0,\left[0,\frac{\psi'^2}{\sqrt{6}},0,0,0,1\right]\right\}\label{e19}
\end{eqnarray}
Let $M$ be a matrix whose columns are the eigenvectors in (\ref{e19}). It can be easily found that $M^{-1}JM$ is a diagonal matrix and  diagonal elements are eigenvalues of $J$, i.e. $J'\equiv M^{-1}JM=diag(-3,-3,-3\gamma,0,0,0)$.
We make a change of variables:
\begin{equation}
\label{e20}
\left[\begin{array}{c}
x'_1\\x'_2\\y'_1\\y'_2\\\lambda'_1\\\lambda'_2
\end{array}\right]
=M^{-1}
\left[\begin{array}{c}
x_1\\x_2\\\bar{y}_1\\\bar{y}_2\\\lambda_1\\\lambda_2
\end{array}\right]
=\left[\begin{array}{c}
x_1-\lambda_1\frac{\psi^2}{\sqrt{6}}\\x_2-\lambda_2\frac{\psi'^2}{\sqrt{6}}\\\bar{y}_1\psi\psi'+\bar{y}_2\psi'^2\\\bar{y}_2\psi^2-\bar{y}_1\psi\psi'\\\lambda_1\\\lambda_2
\end{array}\right]
\end{equation}
Then, we can rewrite the autonomous system in the form of the new variables:
\begin{equation}
\label{e21}
\left[\begin{array}{c}
dx'_1/dN\\dx'_2/dN\\dy'_1/dN\\dy'_2/dN\\d\lambda'_1/dN\\d\lambda'_2/dN
\end{array}\right]
=J'
\left[\begin{array}{c}
x'_1\\x'_2\\y'_1\\y'_2\\\lambda'_1\\\lambda'_2
\end{array}\right]+
\left[\begin{array}{c}
K_1\\K_2\\K_3\\K_4\\K_5\\K_6
\end{array}\right]
\end{equation}
where $K_n=K_i(x'_i,y'_i,\lambda'_i)$, $(i=1,2; n=1,2,3,4,5,6)$. The form of $K_i$ can be easy to get by subsituting the transformations (\ref{e20}) into the equations $dx_i/dN\equiv X_i(x_i,\bar{y}_i,\lambda_i)$, $d\bar{y}_i/dN\equiv \bar{Y}_i(x_i,\bar{y}_i,\lambda_i)$, $d\lambda_i/dN\equiv\Lambda_i(x_i,\bar{y}_i,\lambda_i)$.

According to the center manifold theorem \cite{b1}, there exists a center manifold:
\begin{equation}
\label{e22}
W^c_{loc}=\{(x'_i,y'_i,\lambda'_i)|x'_i\equiv g_i(y'_2,\lambda'_i),y'_1\equiv h(y'_2,\lambda'_i)\}
\end{equation}
and the dynamics of (\ref{e21}) can be restricted to the center manifold.
In (\ref{e22}), $g_i(0,0,0)=h(0,0,0)=J_{g_i}(0,0,0)=J_h(0,0,0)=0$, and $J_{g_i}$, $J_h$ are Jacobi matrix of $g_i$ and $h$ respectively. Therefore one can assume that
\begin{eqnarray}
g_1(y'_2,\lambda'_i)&=&\alpha_1y'^2_2+\alpha_2\lambda'^2_1+\alpha_3\lambda'^2_2+\alpha_4y'_2\lambda'_1+\alpha_5y'_2\lambda'_2+\alpha_6\lambda'_1\lambda'_2+\cdots\label{e23}\\
g_2(y'_2,\lambda'_i)&=&\beta_1y'^2_2+\beta_2\lambda'^2_1+\beta_3\lambda'^2_2+\beta_4y'_2\lambda'_1+\beta_5y'_2\lambda'_2+\beta_6\lambda'_1\lambda'_2+\cdots\label{e24}\\
h(y'_2,\lambda'_i)&=&\delta_1y'^2_2+\delta_2\lambda'^2_1+\delta_3\lambda'^2_2+\delta_4y'_2\lambda'_1+\delta_5y'_2\lambda'_2+\delta_6\lambda'_1\lambda'_2+\cdots\label{e25}
\end{eqnarray}
Because the vector $(dx'_i/dN,dy'_i/dN,d\lambda'_i/dN)$ on a trajectory in $W^c_{loc}$ must tangent to it. This means
\begin{eqnarray}
\frac{dx'_i}{dN}&=&\frac{\partial g_i}{\partial y'_2}\frac{dy'_2}{dN}+\frac{\partial g_i}{\partial \lambda'_1}\frac{d\lambda'_1}{dN}+\frac{\partial g_i}{\partial \lambda'_2}\frac{d\lambda'_2}{dN}\label{e26}\\
\frac{dy'_1}{dN}&=&\frac{\partial h}{\partial y'_2}\frac{dy'_2}{dN}+\frac{\partial h}{\partial \lambda'_1}\frac{d\lambda'_1}{dN}+\frac{\partial h}{\partial \lambda'_2}\frac{d\lambda'_2}{dN}\label{e27}
\end{eqnarray}
Inserting Eqs.(\ref{e23}-\ref{e25}) into Eqs.(\ref{e26},\ref{e27}) we get
\begin{eqnarray}
&& \alpha_n=\left\{
\begin{array}{cc}
0&n\neq 4\\
-\frac{\sqrt{6}}{3}\psi'&n=4
\end{array}
\right.\label{e30}\\
&&\beta_n=\left\{
\begin{array}{cc}
0&n\neq 5\\
-\frac{\sqrt{6}}{3}\psi'&n=5
\end{array}
\right.\label{e28}\\
&&\delta_1=-\frac{\psi'}{2\psi^2},\delta_2=-\frac{\psi^4\psi'}{12},\delta_3=-\frac{\psi'^5}{12},\nonumber\\
&&\delta_4=\delta_5=\delta_6=0\label{e29}
\end{eqnarray}
Then, the dynamics near $A'_2$ is governed by the following equations:
\begin{eqnarray}
\frac{dy'_2}{dN}&=&-\frac{1}{2}\psi^2\psi'(\psi'^2\lambda'^2_2-\psi^2\lambda'^2_1)\label{e31}\\
\frac{d\lambda'_1}{dN}&=&-\psi^2f_1(0)\lambda'^3_1\label{e32}\\
\frac{d\lambda'_2}{dN}&=&-\psi'^2f_2(0)\lambda'^3_2\label{e33}
\end{eqnarray}

Obviously, for any $\psi\in(0,1)$, the component $dy'_2/dN$ of vector $(dx'_i/dN,dy'_i/dN,d\lambda'_i/dN)$ cannot always point to the origin, which means that $A'_2$ is a saddle point even for $f_1(0)>0$ and $f_2(0)>0$.

\begin{figure}[t]
	\includegraphics[width=0.8\textwidth]{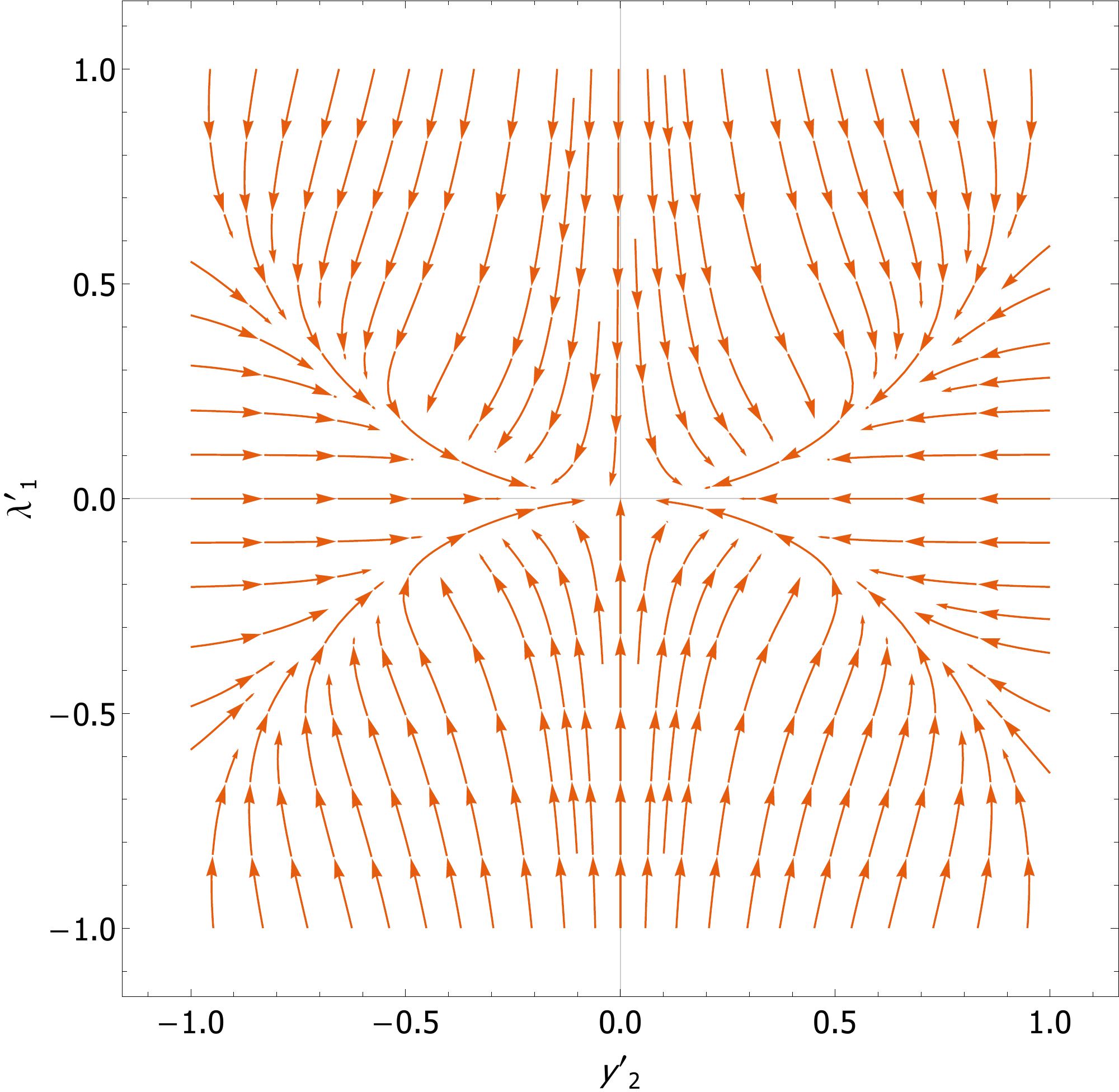}
	\caption{The vector field which determined by Eqs.(\ref{e34},\ref{e35}). We set $\gamma=1$ and $f_1(0)=1$.\label{f1}}
\end{figure}
\begin{figure}[t]
	\includegraphics[width=0.8\textwidth]{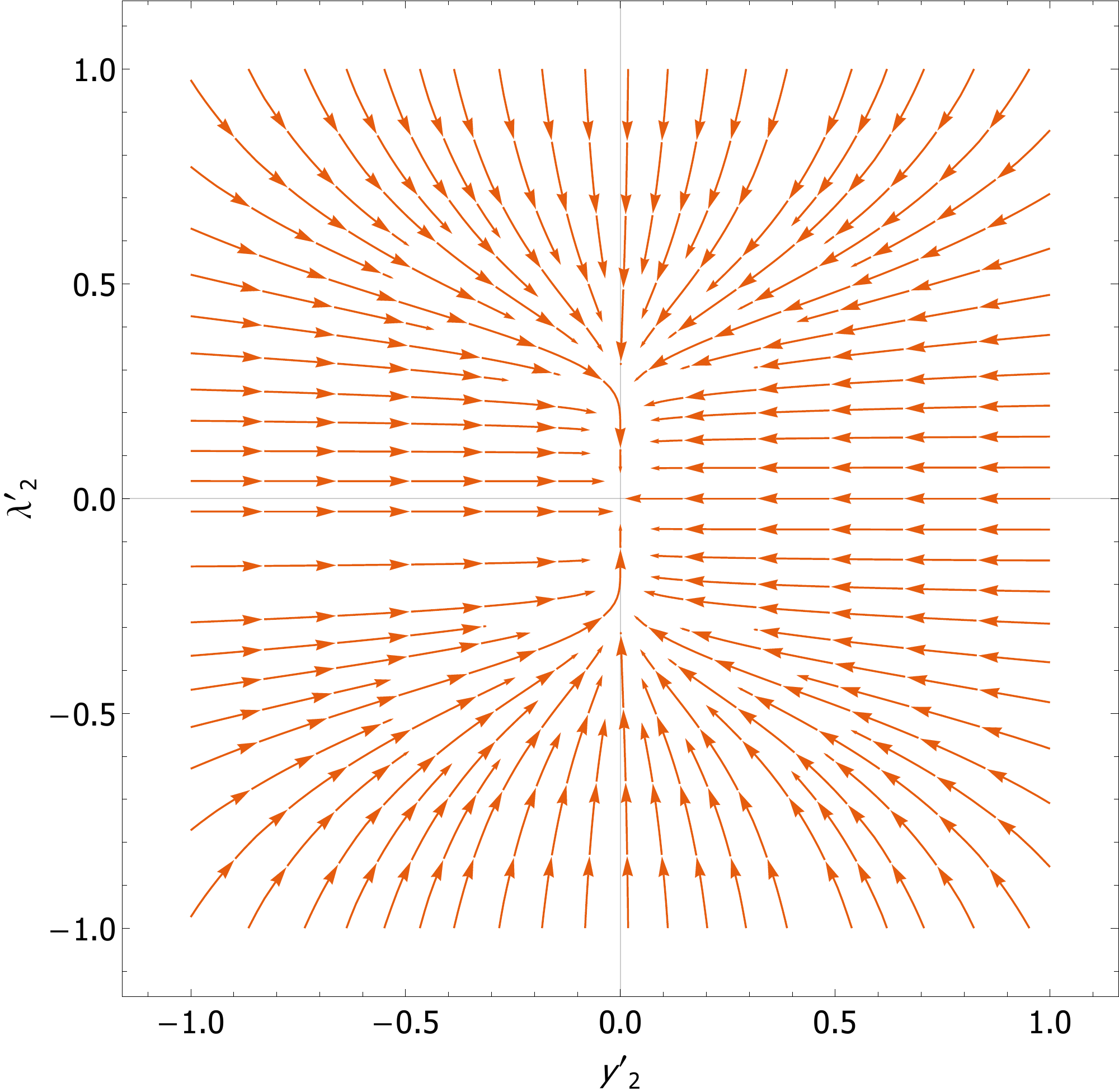}
	\caption{The vector field which determined by Eqs.(\ref{e37},\ref{e39}).We set $\gamma=1$ and $f_2(0)=5$.\label{f3}}
\end{figure}
While, for $\psi=1$ or $\psi=0$, the Jacobi matrix is not the form of Eq.(\ref{e19}). However, we can follow the same method to get the dynamic equations on center manifold. When $\psi=1$, the dynamics near $A'_2$ is governed by the following equations:
\begin{eqnarray}
\frac{dy'_2}{dN}&=&\frac{y'_2\lambda'^2_1}{2}-\frac{3y'^5_2\gamma}{8}-\frac{1}{8}y'^3_2\lambda'^2_1\gamma-\frac{1}{96}y'_2\lambda'^4_1\gamma\label{e34}\\
\frac{d\lambda'_1}{dN}&=&-f_1(0)\lambda'^3_1\label{e35}\\
\frac{d\lambda'_2}{dN}&=&0\label{e36}
\end{eqnarray}
From Eq.(\ref{e36}), we can find that $\lambda'_2$ be a constant near $A'_2$, and the vector field which determined by Eqs.(\ref{e34},\ref{e35}) is shown in Figure \ref{f1}, and one can conclude that $A'_2$ and so $A_2$ is a stabble fixed point when $\psi=1$ and $f_1(0)>0$. We denote this 
fixed point as $A_{2a}$.

However, it shuld be noted that $A_{2a}$ is a \textbf{stabble} point but not a \textbf{asymptotically stabble} point (see \cite{b2} for detail), which means that only
if $|\lambda_2|$ 
 is small enough in the initial moment, then the $A_{2a}$ is the attractor at late time.

On the other hand, when $\psi=0$, the dynamic equations on the center manifold are 
\begin{eqnarray}
\frac{dy'_2}{dN}&=&-3\gamma y'_2\label{e37}\\
\frac{d\lambda'_1}{dN}&=&0\label{e38}\\
\frac{d\lambda'_2}{dN}&=&-f_2(0)\lambda'^3_2\label{e39}
\end{eqnarray}
The vector field which determined by Eqs.(\ref{e37},\ref{e39}) is shown in Figure \ref{f3}
and one can conclude that $A_2$ is also a stabble fixed point when $\psi=0$ and $f_2(0)>0$. We denote this point as $A_{2b}$.

Again, $A_{2b}$ is not a asymptotically stabble point too, and the condition for its attractive behavior is 
that $|\lambda_1|$ is small enough in the initial moment.

It is found that, there are two stabble fixed points: $A_{2a}$ and $A_{2b}$ for $\psi=1$ 
and $\psi=0$ when $f_1(0)>0$ and $f_2(0)>0$ respectively.

It is worth noting that, these two stabble points are both single scalar field model. This result is contrary to 
the conclusion of \cite{r5}, in which the author obtain the rather surprising result that a single scalar field model \textbf{never} dominates at late times if one consider the exponential potential.

If the potential of one scalar field satisfy the condition of  $f(0)>0$ and the other scalar field satisfy the condition of $|\lambda|\ll 1$(i.e. slowly varying potential), then the universe with two scalar fields will enter the single scalar field era at late time, and the scalar field whoes potential slowly varying at initial moment will decay to zero.

\begin{figure}[t]
	\includegraphics[width=0.8\textwidth]{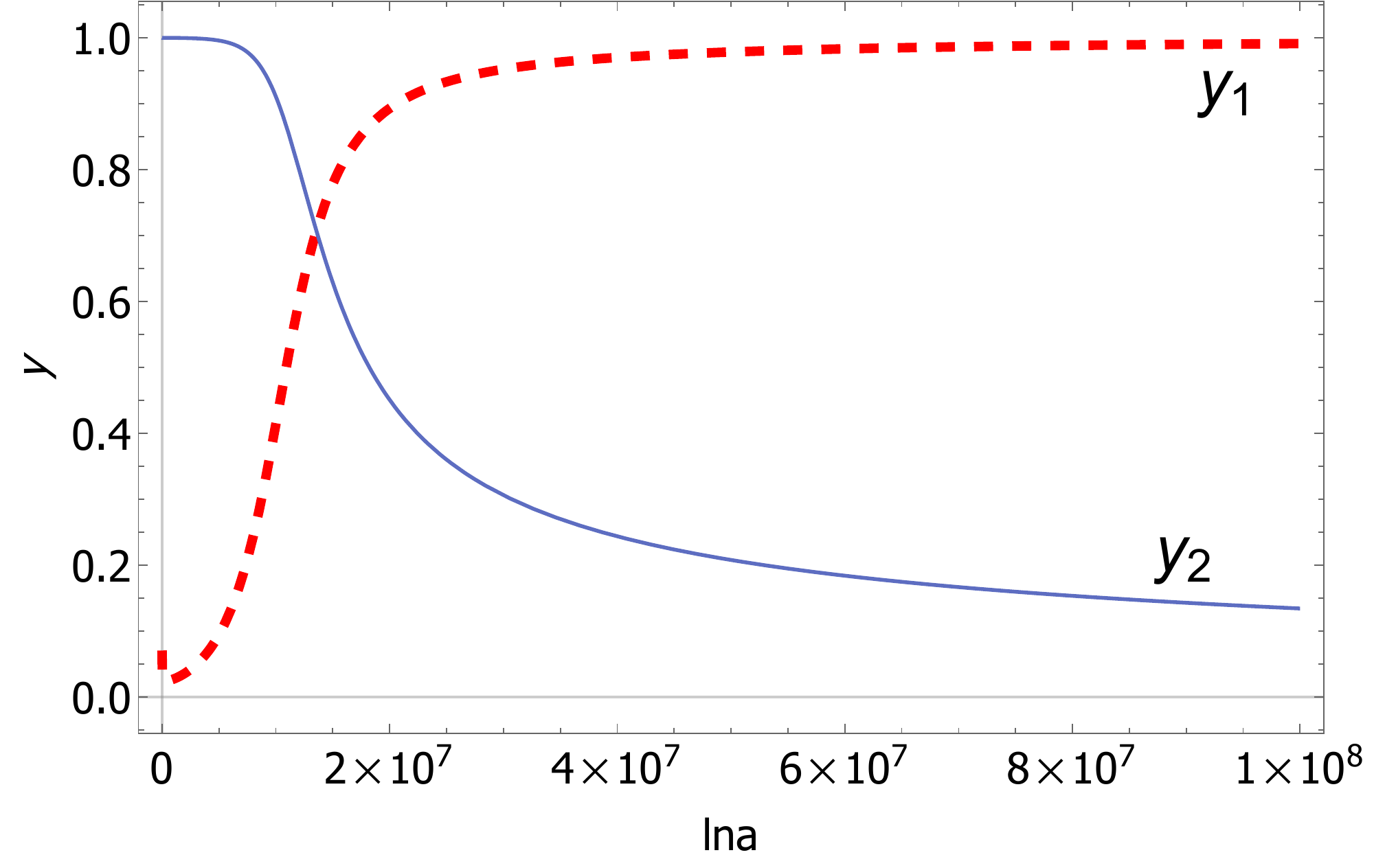}
	\caption{ The number result for evolution of $y_1$ and $y_2$.\label{f2}}
\end{figure}

In Fig.\ref{f2}, we give the number result for evolution of $y_1$ and $y_2$, in which we set $f_1(\lambda)=1$ (i.e. the potential $V_1(\phi)=\frac{\mathcal{C}_1}{\phi+\mathcal{C}_2}$, $\mathcal{C}_1,\mathcal{C}_2$ are constant of integration.) and $f_2(0)=1$.
The initial conditions we have chosen are $x_1(0)=x_2(0)=1,y_1(0)=0.1,y_2(0)=1,\lambda_1(0)=1,\lambda_2(0)=0.001$. We can see that, $y_1$ will dominates the energy density at late time and $y_2$ will decay to zero even if $y_2\gg y_1$ at initial moment.

\section{Mutiple Attractors Case\label{s4} }
Other interesting case is when $f_1(0)>0$ and $f_2(0)>0$ are satisfied at the same time, then there will 
be two late time attractors. Which scalar field dominate the universe at late time depend on the  initial condition.

The eigenvalues of the Jacobi matrix for the assisted scalar field dominated
solution $F_6$ are
\begin{eqnarray}
&&\left\{\lambda_{eff}^2-3\gamma,\frac{\lambda_{eff}^2-6}{2}, -\frac{1}{4}\left[(6-\lambda_{eff}^2)\pm\sqrt{(6-\lambda_{eff}^2)-8\lambda_{eff}^2(6-\lambda_{eff}^2)}\right]\right.\nonumber\\
&&\left.-\lambda_1^*\lambda_{eff}^2f_1'(\lambda_1^*),-\lambda_2^*\lambda_{eff}^2f_2'(\lambda_2^*)\right\}\label{e17}
\end{eqnarray}
where $f_1'(\lambda)\equiv df_1/d\lambda$, $f_2'(\lambda)\equiv df_2/d\lambda$.

Because the condition for existence of $F_6$ is $|\lambda_{eff}|<\sqrt{6}$, therefore the sufficient conditions for $F_6$ is a stabble point are
\begin{equation}
\label{e18}
\lambda_{eff}^2<3\gamma,~~\lambda_1^*f_1'(\lambda_1^*)>0,~~\lambda_2^*f_2'(\lambda_2^*)>0
\end{equation}

On the other hand, the eigenvalues of the assisted scaling solution $F_7$ is 
\begin{eqnarray}
&&\left\{-\frac{3\left[(2-\gamma)\pm\sqrt{(2-\gamma)^2-8\gamma(2-\gamma)}\right]}{4},\right.\nonumber\\
&&-\frac{3\left[(2-\gamma)\pm\frac{1}{\lambda_{eff}}\sqrt{(2-\gamma)[24\gamma^2-(9\gamma-2)\lambda^2_{eff}]}\right]}{4},\nonumber\\
&&\left.-3\lambda^*_1f'_1(\lambda_1^*)\gamma,-3\lambda^*_2f'_2(\lambda_2^*)\gamma\right\}\label{e40}
\end{eqnarray}
The necessary condition for existence of $F_7$ is $\lambda_{eff}^2>3\gamma$, and the conditions for stabble of $F_7$ are $\lambda_1^*f_1'(\lambda_1^*)>0,\lambda_2^*f_2'(\lambda_2^*)>0$.

The scaling solution is useful to solve the coincidence problem of dark energy.
If there is only one value of $\lambda$ to make $f_i(\lambda)=0$, i.e. there is only one zero point for functions $f_i$, the assisted scalar field dominated solution $F_6$ 
and the assisted scaling solution $F_7$ cannot  exist and be stabble at same time.
This means that, the universe cannot exit from the scaling regime and evolve to recent acceleration era. 
One method to realize the evolve from a scaling attractor to a acceleration attractor, in which the potential can be approximated to two different potentials when $\phi$ evolves into different ranges \cite{r6,r7,r8}.

There is another way to solve this issue base on multiple attractors. In \cite{r3}, 
the author proposed a scenario of universe which could evolve from a scaling solution to a de-Sitter-like attractor by introducing a field whose value changed by a certain amount in a short time. From discussion above, one can find that scaling solution $F_7$ and de-Sitter-like attractor $A_{2a}$ or $A_{2b}$ can also be stabble at the same time. 

On the other hand, if there are multiple zero points for  $f_i$, the solution of multiple attractors can also be found. To discuss this issue, we assume $f_1(\lambda)$ has two different zero points $\lambda^*_{1a}$ and $\lambda^*_{1b}$, $f_2(\lambda)$ has one zero point $\lambda^*_2$.
Under these assumptions, there are two scalar field dominated solutions $F_{6a}$ and $F_{6b}$ and two scaling solutions $F_{7a}$ and $F_{7b}$ corresponding to
$\lambda^*_{1a}$ and $\lambda^*_{1b}$ respectively. 

It is worth noting that two scalling solutions are problematic \cite{r9}. In order to avoid this situation we also assume 
\begin{equation}
\label{e41}
\left(\frac{1}{\lambda^{*2}_{1a}}+\frac{1}{\lambda^{*2}_{2}}\right)^{-2}<3\gamma<\left(\frac{1}{\lambda^{*2}_{1b}}+\frac{1}{\lambda^{*2}_{2}}\right)^{-2}
\end{equation}
Which means that the scaling solution $F_{7b}$ exist but $F_{7a}$ does not. On the other hand, if Eq.(\ref{e41}) is satisfied, the scalar  field dominated solution $F_{6a}$ is stabble but $F_{6b}$ is not.

Moreover, to make $F_{6a}$ and $F_{7b}$ are stabble at the same time, $\lambda_{1a}^*f_1'(\lambda_{1a}^*)>0,\lambda_{1b}^*f_1'(\lambda_{1b}^*)>0$ must satisfied simultaneously.  Generally speaking, if $f_1(\lambda)$ has two zero points, then one of  $f_1'(\lambda_{1a}^*)$ 
and $f_1'(\lambda_{1b}^*)$ must be positive and the other must be negative, which means that 
\begin{equation}
\label{e42}
\lambda_{1a}^*\lambda_{1b}^*<0
\end{equation}
\section{SUMMARY\label{s5}}
In summary, we discuss the dynamics of two-scalar-field cosmologyical models. Unlike in situation of 
exponential potential, we found the single scalar field dominated late time attractors. The universe with two scalar fields will enter the single scalar field era at late time, and the scalar field whoes potential varying  slowly at initial moment will decay to zero. 

Multiple attractors model is useful to realize the 
evolve from a scaling era to recent acceleration era. In the models beyond exponential potential, the 
multiple attractors can be obtain easily. We find that,  scaling solution $F_7$ and de-Sitter-like attractor $A_{2a}$ or $A_{2b}$ can also be stabble at the same time as in\cite{r3}.  This provides a possibility that the 
universe evolve from a assisted scaling era to a single scalar field acceleration era.

Moreover, if $f_i$ has multiple zero points, we also can construct a model with assisted scaling solution and assisted scalar field dominated solution. We give the conditions Eqs.(\ref{e41},\ref{e42}) for multiple
attractors models.

\section*{Acknowledgments}

We thank Hong Gao for useful discussions. This work was supported by the National Natural Science Foundation of China under Grant No.11447198 (Fund of theoretical physics) and
the Fundamental Research Funds for the Central Universities of Ministry of Education of China under Grants No. 3132017064.

%\begin{thebibliography}{000} %for 3 digits
%\begin{thebibliography}{00}  %for 2 digits
\bibliographystyle{ws-ijmpd}
\bibliography{ws-ijmpd}

\end{document}